# Gradient spin echo enhanced proton precession magnetometer: a novel system for field gradient measurement

Farrokh Sarreshtedari*, Faeze Mahboubian, Mohammad Hadi Sardari
Magnetic Resonance Research Laboratory, Department of Physics, College of Science, University of Tehran, 1439955961, Tehran, Iran.

*Corresponding author (f.sarreshtedari@ut.ac.ir)

## Abstract
Gradient spin echo enhanced proton precession magnetometer is a novel system which can measure the first order gradient of the background field in addition to the magnetic field. The system includes a conventional proton precession magnetometer equipped with Maxwell coil pair and electronics which allow to conduct the gradient spin echo experiment. In gradient spin echo process, based on the background gradient field, the switching gradient field and the switching reversal time, the spin echo signal forms at a theoretically predictable time. The important advantage of this approach is that in contrast to conventional proton gradiometers which measure the magnetic field difference between two different points, gradient spin echo enhanced proton magnetometer measures the field gradient at the same position where the magnetic field is being measured. It is shown that by using this system the background gradient field is measured with an average root mean square error of $0.02\mu T/m$ for gradient fields in the range of $-0.25\mu T/m$ to $+0.25\mu T/m$. By optimization of this system, the mentioned error could be significantly decreased and the instrument could be used for many different applications.

## Introduction
Proton precession magnetometer is one of the most widely used instruments for earth field measurements and geomagnetic surveys [1-3]. Although the introduction of these scalar magnetometers dates back to more than half a century ago, because of their high resolution (about 0.01 nT), high stability and simplicity, they are still popular and different research groups work on methods for increasing the detection sensitivity [4-8]. As many magnetic survey applications require the measuring of the field gradient [9], proton gradiometers are also widely used by simultaneous reading of two magnetometers which are placed at a distance from each other, a so-called baseline. These gradiometers do not measure the real field gradient at a specific point and instead measure the magnetic field difference between two points. Because of that they could be only useful in environments with low field gradient variations. Proton precession magnetometers use the principle of Nuclear Magnetic Resonance (NMR) for measuring the absolute value of the magnetic field [1]. In contrast to magnetic resonance imaging (MRI) or NMR spectrometers which incorporate $\pi/2$-RF pulses for generating the transverse magnetization, in proton precession magnetometers, a polarizing field is responsible for this excitation. When the polarizing field is switched off, the coils receive the free induction decay (FID) signal in the transverse plane due to precession of transverse magnetization vector at Larmor frequency. Due to the dephasing of the ensemble, the transverse magnetization and the envelope of the FID signal exponentially decay. When the magnetic field is homogeneous in the sample volume, the decay time constant is described by spin-spin relaxation time ($T_2$). However, the presence of the magnetic field gradient within the sample volume, causes the inhomogeneous broadening and quadratically degrades the dephasing [10, 11]. In this case, the dephasing time constant is described by effective spin-spin coupling ($T_2^*$). Spin echo experiment is a standard method for $T_2$ measurement in magnetic resonance imaging (MRI) and NMR spectrometers. In this experiment $T_2$ could be obtained regardless of the external gradient or other inhomogeneous broadening sources. Spin echo pulse sequence includes a $\pi/2$-pulse followed by a $\pi$-pulse after a delay of $\tau$. In this sequence, the echo signal would be generated when a $\tau$ interval is passed after second RF pulse. In reference [12], incorporating a SQUID based NMR system, the spin echo sequence is used to overcome the inhomogeneous broadening which increased the SNR of the magnetic measurement. Here, we have demonstrated a proton precession magnetometer which is equipped with a gradient spin echo setup. This proposed setup not only preserves the simplicity of the proton magnetometer, but also provides the ability of gradient field measurement for this magnetometer which can significantly improves its applications. The gradient field measurement by gradient spin echo enhanced proton precession magnetometer is accomplished at the same position of the sensor. As a result, this novel system measures the real field gradient at the proton reservoir which makes it useful for measurement in environments with strong spatial variation of the field gradient. Furthermore, it is worth mentioning that the novel introduced method for gradient measurement based on time shift of gradient echo response could be also developed for other magnetometers incorporating magnetic resonance phenomena.

## Gradient spin echo for gradient measurement
Gradient spin echo (GRE) is one of the important sequences in magnetic resonance imaging (MRI) and NMR spectrometers. In these systems, the GRE is produced by a $\pi/2$-RF pulse in conjunction with a magnetic field gradient reversal [10, 13]. Fig.1 shows the schematics of the GRE sequence which include the dephasing and rephrasing intervals. In Fig. 1(a) a common NMR signal with effective decay time constant $T_2^*$ is depicted. Fig. 1(b) schematically shows that when an externally gradient field is applied by start of NMR signal production and reversed at $\tau_1$, the spins refocus and the echo is generated after a time interval of $\tau_2$. The key point is that the gradient spin echo forms when the integral of the field gradient over time is equal to zero [10].

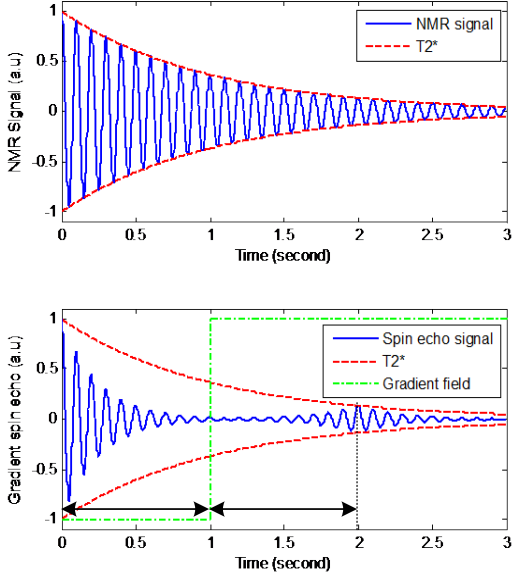

Fig 1. The schematics of the gradient spin echo sequence (a) NMR signal without echo (b) Gradient spin echo is obtained at T=2s when the gradient field reversal is taken place at T=1s.

If the background field gradient is zero, the interval between the gradient reversal and the echo ($\tau_2$) is the same as the interval between the start of the NMR signal production and the gradient reversal ($\tau_1$). However, if a background field gradient exists, based on its direction, the echo forms sooner or later ($\tau_1 \neq \tau_2$). In this regard, the difference between $\tau_1$ and $\tau_2$ depends on the background gradient field and its direction. If the background field gradient is $G_{background}$ and the used gradient field for echo experiment is $G_{echo}$, in the interval $\tau_1$ the total gradient field is $G_{background} + G_{echo}$ while in the interval $\tau_2$ it is $G_{background} - G_{echo}$. Since the integral of the field gradient is zero at $\tau_2$, $G_{background}$ could be found using the equation (1).

$$G_{background} = G_{echo}\left(\tau_2 - \tau_1\right)/\left(\tau_2 + \tau_1\right) \qquad (1)$$

It should be noted that the reversal of the gradient field refocuses only those spins that are dephased by action of the applied gradient itself and so the $T_2^*$ is not influenced by the applied gradient field. This is shown in Fig.1 (b), when the amplitude of the obtained echo reaches to the $T_2^*$ envelope. This is why the ratio of the echo to initial NMR signal could be used for determining the $T_2^*$. Although the quadratically dependence of the $T_2^*$ to the field gradient gives a measure of the gradient, it could not be used as a measurement method. However in this work it is shown that measuring the difference between $\tau_1$ and $\tau_2$ in GRE process is a very useful approach for measurement of different field gradient components.

## Experimental setup

Figure 2 shows the block diagram of the implemented proton magnetometer which has the feature of conducting the gradient spin echo experiment. The system includes an implemented proton magnetometer [14] which its electronics is modified for providing a reversable current to a Maxwell coil setup which produce the homogeneous gradient field. Maxwell coil pair is a configuration to generate magnetic field gradient using two coils carrying the same current in opposite directions. These coils have the same radius R and the distance between them is equal to $\sqrt{3}$R. This configuration produces an approximately uniform magnetic field gradient in the center of the coils. In our experimental setup the Maxwell coils are designed in such a way that the field gradient error at the sides of the sensor reservoir is below 1%. For achieving this accuracy, the radius of the constructed Maxwell coils are R=65cm and the distance between the coils is d=113cm. In this system, the provided gradient field for an applied current (I) is $\left(\partial B/\partial r\right)/I = 44\,\mu T/m.A$. For investigation of the environment gradient field effect on the gradient echo experiment, we have also wounded a pair of new Maxwell coils on the same structure for applying the background gradient field. We call these sets of coils as background gradient coils in this paper.

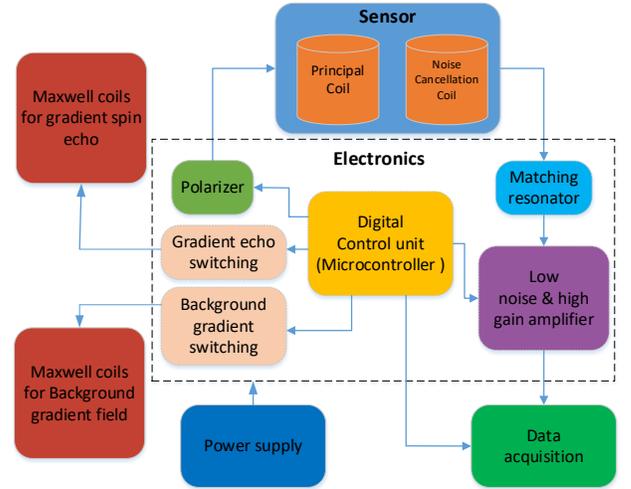

Fig 2. Block diagram of the GRE enhanced proton precession magnetometer.

The head of the implemented proton magnetometer consists of a principal coil which is wound on a cylindrical reservoir with a diameter of 8cm and length of 15cm. The principal coil is beside an identical coil with the same dimension and number of turns for noise cancellation. The sensor reservoir is filled with a solution of water and ethanol (5:1). For environmental noise cancellation, the two coils are wound in opposite direction and connected in series. In such configuration, a same induced voltage on both coils would be subtracted from each other and cancelled out. The principle of magnetometer operation is based on polarization and detection phases [2, 14]. The polarizer generates a magnetic field in the reservoir which is much stronger than the earth magnetic field. This applied field polarizes the proton spins along the coil axis by spin-lattice time constant ($T_1$). The polarization period starts by connecting a dedicated battery source to the principal coil via an electronic relay and a HEXFET transistor. The transistor is responsible for immediate reduction of the coil current to zero at the end of polarization period [2]. When the entire stored energy in the coil is dissipated in the HEXFET and the dc resistance of the coil, the electronic relay connects the sensor coil to a parallel capacitor bank and the electronic front end amplifier. The coil inductance and the matched capacitors comprise a LC resonator tuned on the earth field Larmor frequency for enhancing the currents in the coil by electromagnetic resonance [10]. The sensor induced precession signal is a few microvolts which is increased to about one volt in the low noise amplifier unit. The amplifier include a copper shielded front end low noise differential amplifier followed by successive filtering and amplification units. Due to the very high gain characteristics of the amplifier (about 150000), electromagnetic compatibility (EMC) considerations like proper grounding, shielding and filtering is required at the system and board levels for preventing the unwanted oscillation. Furthermore, for maximum noise reduction, the bandwidth of the amplifier is limited to a few kHz around the center frequency of 2.1 kHz. The amplified signal is then acquired by a 24-bit, 200kS/s, USB data acquisition module. As shown in Fig. 2, the timing sequence for polarization, detection and data acquisition is controlled by a microcontroller based digital unit. In GRE enhanced proton precession magnetometer, the digital unit also controls the GRE timing process. In this system the user can control the GRE parameters including the intensity of the applied gradient field and the time duration before field reversing ($\tau_1$). The $\tau_1$ interval could be also swept for acquiring two dimensional GRE sequences [10]. Figure 3 shows the experimental setup of our system, including the proton precession magnetometer and the Maxwell coils. The magnetometer coil is placed at the center of the Maxwell setup in such a way that their axes are the same. The whole system is placed in an environment with homogeneous background magnetic field.

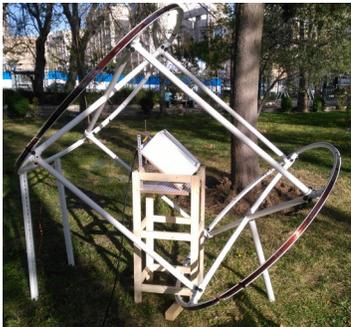

Fig 3. The experimental setup of the GRE enhanced proton magnetometer.

If we choose the x-axis of the coordinate system along the direction of the sensor coil, the Maxwell setup apply a gradient component of $dBx/dx$. This is the same component which is being measured by the current GRE enhanced proton magnetometer. The other field gradient components are not influenced by the Maxwell pair and although they have their own effect on $T_2^*$ of the NMR signal but have no effect on the spin echo signal. It should be noted that by utilizing more advanced gradient coil configurations instead of simple

Maxwell setup, different field gradient components could be measured by GRE enhanced proton magnetometer.

## Results and discussion

Fig. 4 shows the gradient spin echo experiment result when the environment gradient field is approximately zero. Fig. 4(a) shows the proton precession FID signal without echo experiment. Fig. 4(b) shows the FID signal when the used gradient field for GRE experiment is applied but no gradient reversal takes place. In Fig. 4(c)-(e) the gradient reversal switching for echo experiment ($\tau_1$) is increased from 0.3s to 0.5s by steps of 0.1s. In each part, the result of the echo experiment is overlapping with the FID signal of case (a) for emphasizing the non-changing $T_2^*$ during spin echo experiments.

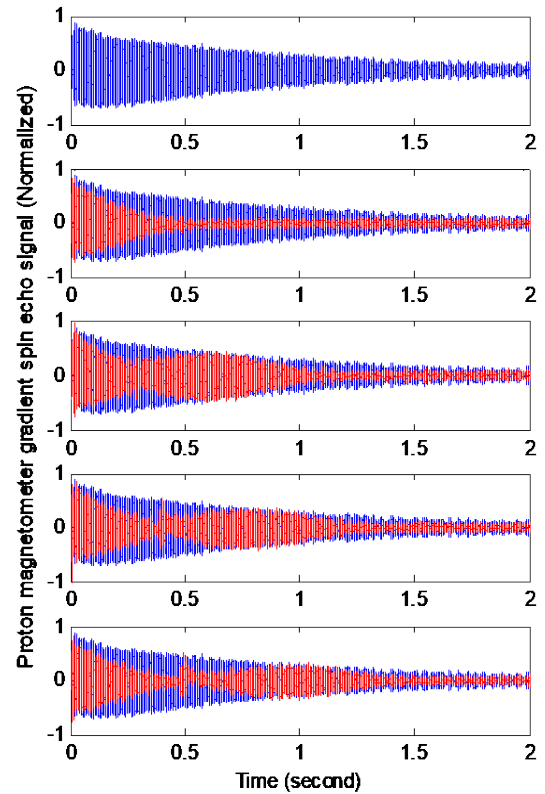

Fig 4. The gradient spin echo experimental results in an environment with zero background gradient using the proposed proton magnetometer. (a) The FID signal without echo experiment, (b) The FID signal when the gradient field without reversal is applied, (c) Gradient spin echo with $\tau_1$=0.3s, (d) Gradient spin echo with $\tau_1$=0.4s, (e) Gradient spin echo with $\tau_1$=0.5s. In this experiments, the amplitude of the switching gradient field is 0.91μT/m and the FID signal of (a) is underlying in graphs (b)-(e).

It is evident that as the environmental gradient field is zero, the echo formation time ($\tau_2$) is equal to gradient switching time ($\tau_1$) in all parts of Fig.4 (c)-(e). Furthermore, in each part of this figure, the echo signal reaches to the $T_2^*$ of the no gradient case, as the dephasing of the spins because of the applied gradient field is refocused at echo time. This was previously explained in schematics of Fig. 1(b). It should be noted that a low pass filter post processing is being used to the raw data of this figure for high frequency noise cancellation. For quantitative investigation of the background gradient effect on the spin echo result, this experiment is accomplished in the presence of determined background gradient fields. Fig. 5(a)-(f) shows the gradient spin echo result when different background gradient fields are applied to the magnetometer during the echo experiment. In Fig. 5(a)-(e), the applied gradient fields are +0.21μT/m, +0.14μT/m, 0.00μT/m, -0.14μT/m, -0.17μT/m respectively. The switching gradient field is +0.91μT/m and the switching reversal time ($\tau_1$) is 0.3s for all the measurements.

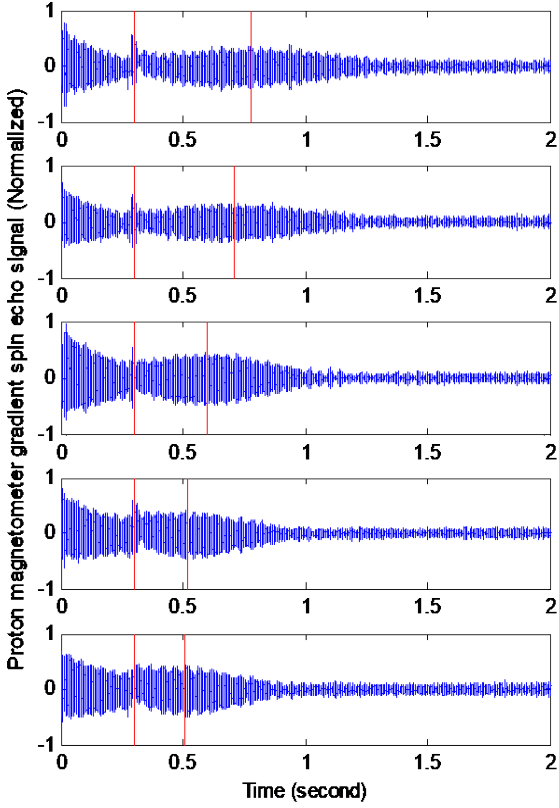

Fig 5. The gradient spin echo experimental results for different background gradient. The red lines indicate the reversal gradient time of 0.3s and the theoretical calculated echo time, (a) The background gradient field is +0.21μT/m, (b) The background gradient field is +0.14μT/m, (c) The background gradient field is 0.00μT/m, (d) The background gradient field is -0.14μT/m, (e) The background gradient field is -0.17μT/m.

The interval $\tau_1$ is 0.3s for all experiments shown in Figs.5. The obtained $\tau_2$ is different for each experiment. The red lines in Fig.5 from left to right indicate the reversal gradient time 0.3s and the theoretical calculated echo time using equation (1), respectively. A very good agreement can be observed between the calculated $\tau_2$ and the experimental echo time result. It is clear that for the cases with positive background gradient, $\tau_2$ is greater than $\tau_1$. However, in the cases which the background gradient is negative, $\tau_2$ is less than $\tau_1$. It should be noted that the background gradient is positive when it is in the same direction of the initial echo gradient field (before reversal). In this regard, by knowing the gradient field for echo experiment and also the considered reversal switching time of $\tau_1$, and by measuring $\tau_2$ we can obtain the background gradient field. It should be noted that for precise measurement of background gradient field, one can conducts two successive experiments with same $\tau_1$ interval and opposite gradient echo field polarity.

In this situation, the $\tau_2$ is greater than $\tau_1$ for one experiment and less than $\tau_1$ for the other. By averaging the gradient measurement of the two experiments, more accurate result could be obtained. As previously discussed, the $T_2^*$ could be also found by measuring the relative amplitude of the echo signal. All the experimental results shown in Figs. 4-6 have been used a fixed and same switching gradient field +0.91μT/m for obtaining the spin echo. The value of this gradient field is also a key parameter for precise background gradient measurements. Fig. 6 shows the gradient spin echo experimental results with two different switching gradient fields 0.91μT/m and 1.07μT/m and also two background gradient fields +0.25μT/m and -0.25μT/m. The switching reversal time ($\tau_1$) is 0.4s for all the cases. The red lines in Fig.6 indicate the reversal gradient time 0.4s and the theoretical calculated echo time.

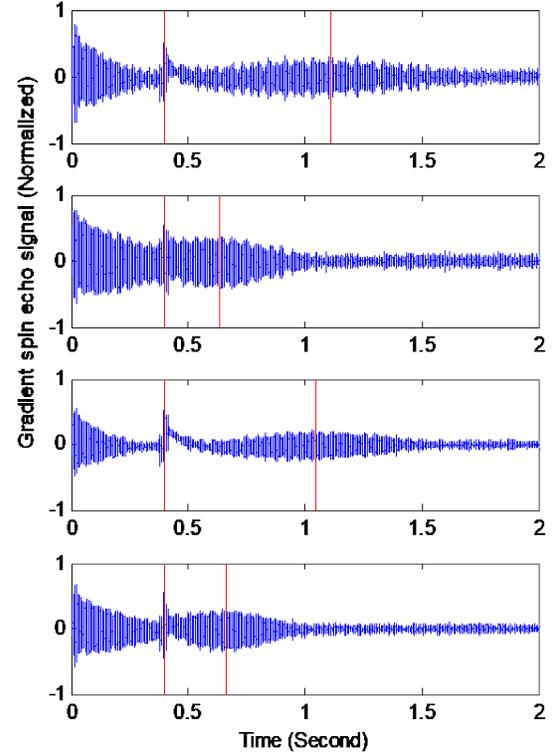

Fig. 6. The gradient spin echo experimental results for two different switching gradient fields and also two background gradient fields. The red lines indicate the reversal gradient time 0.4s and the theoretical calculated echo time, (a) The switching gradient is 0.91μT/m and the background gradient field is +0.25μT/m, (b) The switching gradient is 0.91μT/m and the background gradient field is -0.25μT/m, (c) The switching gradient is 1.07μT /m and the background gradient field is +0.25μT/m, (d) The switching gradient is 1.07μT/m and the background gradient field is -0.25μT/m,

Fig.6 shows that the larger switching gradient field (figures 6(c) and 6(d)) results in the greater damping rate of the FID and echo signal. Furthermore, in figures 6(a) and 6(c) the background gradient and the switching gradient have the same signs in the interval 0 to 0.4s and have different signs in the echo part of the signals. The mentioned signs are vice versa in figures 6(b) and 6(d). Hence the damping rate in the interval 0 to 0.4s for figures 6(a) and 6(c) is greater than the damping rate of figures 6(b) and 6(d), while the damping rate in echo part of figures 6(a) and 6(c) is less than the damping rate in figures 6(b) and 6(d). Considering the fact that the echo signal exponentially decays and signal to noise ratio decreases with time, measuring the large $\tau_2$ values is more difficult than small values. Since choosing the appropriate switching gradient field is important for accurate background gradient measurement. Fig. 7 shows the obtained results for measuring different applied background gradient fields using the gradient echo experiment with different reversal switching time of $\tau_1$.

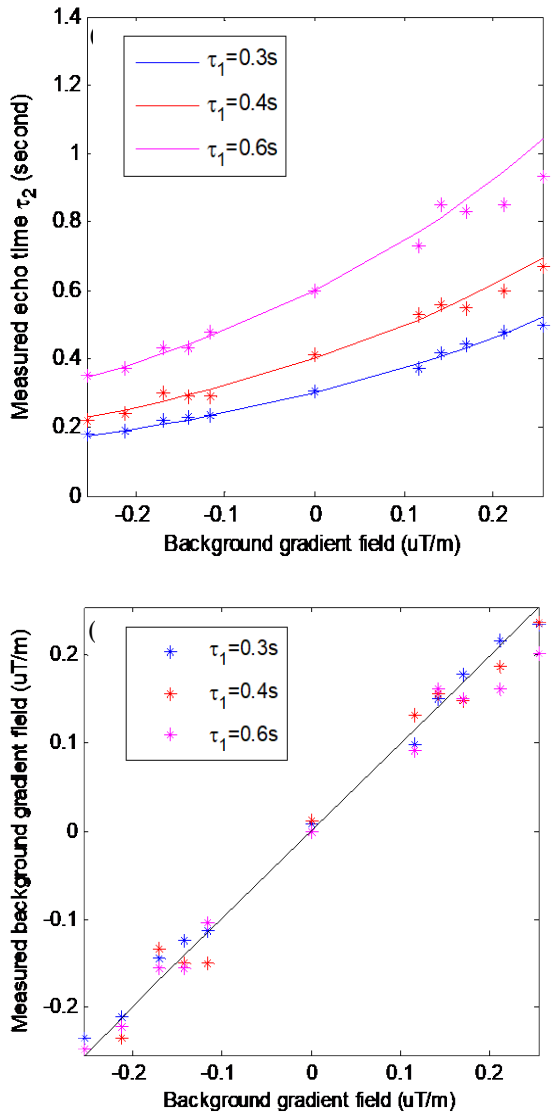

Fig 7. The results of measuring background gradient field using gradient spin echo experiments with different $\tau_1$ values. The solid lines are the theoretical calculated results and the markers are the measurement results. The switching gradient field is $+0.91\mu T/m$ for all measurements. (a) The $\tau_2$ measurement results in gradient spin echo experiments, (b) The calculated values of background gradient field using the measured $\tau_2$ shown in previous part.

Fig. 7(a) shows the measured $\tau_2$ values and Fig. 7(b) shows the obtained values for background gradient field. For each points of Fig. 7(b), using equation (1) with the measured $\tau_2$ values and the known switching gradient field ($G_{echo}= +0.91\mu T/m$), the background gradient field ($G_{background}$) is calculated. The achieved root mean square error of gradient measurements in Fig. 7(b) for $\tau_1$=0.3s and $\tau_1$=0.4s and $\tau_1$=0.6s are 0.018$\mu T/m$, 0.022$\mu T/m$ and 0.036$\mu T/m$ respectively. It is worth mentioning that different factors determine the accuracy of the background gradient measurement using this proposed system. Some these factors are the signal to noise ratio of the proton precession FID signal, the signal processing algorithm for $\tau_2$ measurement from raw data, the choice of switching gradient field ($G_{echo}$) and the switching reversal time ($\tau_1$), the homogeneity of the background gradient and the alignment and accuracy of the Maxwell coils for producing gradient field inside the proton precession magnetometer.

## Conclusion

In this work we have proposed the incorporation of gradient spin echo process in proton precession magnetometer for background gradient field measurement. It is shown that by modification of the proton magnetometer control unit for applying appropriate switching gradient field and also using Maxwell coils for homogeneous gradient generation, the conventional proton magnetometer could be used for first order gradient measurement. Different spin echo experiments are conducted for evaluation of this method for gradient measurement. The experiments are accomplished by using different switching gradient fields, different gradient reversal times and also at different background fields. A very good agreement is achieved in the comparison of the measured $\tau_2$ values and the theoretical predictions. The obtained average root mean square error for gradient measurement is $0.02\mu T/m$ which could be substantially decreased by noise cancellation and using more appropriate signal processing techniques for measuring the echo time from raw FID signals. We believe that improving this system by using multi-dimensional gradient coil setup, decreasing the system volume, developing adequate signal processing algorithms and optimizing the gradient spin echo process by proper choice of switching gradient field and reversal time, makes it a useful instrument for different geophysical applications.

## Acknowledgment

The authors would like to thanks Iran National Science Foundation (INSF) for their support.

## Data availability

The data that support the findings of this study are available from the corresponding author upon reasonable request.

## References

[1] P. Ripka, "Magnetic Sensors and Magnetometer", Artech House Publication, 2001.

[2] J.A. Koehler, "Proton Precession Magnetometers", Rev. 3 Comox, BC, Canada, 52 pp., 2012.

[3] L. J. Cahill, Jr., and J. A. van Allen, "High altitude measurements of the earth's magnetic field with a proton precession magnetometer," J. Geophys. Res., vol. 61, no. 3, pp. 547–558, Sep. 1956.

[4] H. Liu, H. Dong, J. Ge, Z. Liu, Z. Yuan, J. Zhu, H. Zhang, "High-Precision Sensor Tuning of Proton Precession Magnetometer by Combining Principal Component Analysis and Singular Value Decomposition," in IEEE Sensors Journal, vol. 19, no. 21, pp. 9688-9696, 1 Nov.1, 2019.

[5] H. Liua, H. Dong, Z. Liu, J. Ge, B. Bai and C. Zhang, "Noise characterization for the FID signal from proton precession magnetometer", Journal of Instrumentation, Volume 12, 2017.

[6] H. Dong, H. Liu, J. Ge, Z. Yuan, and Z. Zhao, "A High-Precision Frequency Measurement Algorithm for FID Signal of Proton Magnetometer", IEEE Trans. Instrum. Meas., V. 65, N. 4, 2016.

[7] A. Y Denisov, V. A Sapunov, B. Rubinstein, "Broadband mode in proton-precession magnetometers with signal processing regression methods", Measurement Science and Technology, Volume 25, Number 5, 2014.

[8] A. Y. Denisov, O. V. Denisova, V. A. Sapunov, and S. Y. Khomutov, "Measurement quality estimation of proton-precession magnetometers", Earth Planets Space, 58, 707–710, 2006.

[9] C. Harrison, J. Southam, "Magnetic Field Gradients and Their Uses in the Study of the Earth's Magnetic Field", Journal of geomagnetism and geoelectricity. 43. 10, pp585-599, 1991.

[10] M. H. Levitt, "Spin Dynamics: Basics of Nuclear Magnetic Resonance", 2nd ed, John Wiley & Sons Ltd, 2008.

[11] C. Gemmel, W. Heil, S. Karpuk, K. Lenz, Ch. Ludwig, Yu. Sobolev, K. Tullney, M. Burghoff, W. Kilian, S. Knappe-Grüneberg, W. Müller, A. Schnabel, F. Seifert, L. Trahms, St. Baeßler, "Ultra-sensitive magnetometry based on free precession of nuclear spins", Eur. Phys. J. D 57 303–320, 2010.

[12] J. H. Shim, S. Lee, S. Hwang, K. K. Yu, K. Kim, "Proton spin-echo magnetometer: a novel approach for magnetic field measurement in residual field gradient", Metrologia 52 496–501, 2015.

[13] Elster AD. Gradient echo imaging: techniques and acronyms. Radiology 1993; 186:1-8.

[14] F. Mahboubian, H, Sardari, S. Sadeghi, F. Sarreshtedari; "Design and Implementation of a Low Noise Earth Field Proton Precession Magnetometer", 27rd Iranian Conference on Electrical Engineering (ICEE2019), Yazd, Iran, 2019.